\documentclass[twocolumn,a4paper,superscriptaddress,floatfix,showpacs]{revtex4}
\usepackage{amsmath,amssymb,epsfig,color,textcase}

\begin{document}

\title{Electronic structure and magnetic properties of pyroxenes 
(Li,Na)TM(Si,Ge)$_2$O$_6$: Low-dimensional magnets with 90$^0$ bonds}
\author{S.V.~Streltsov}
\affiliation{Institute of Metal Physics, S.Kovalevskoy St. 18, 620041 Ekaterinburg GSP-170, Russia}
\affiliation{Ural State Technical University, Mira St. 19, 620002 Ekaterinburg, Russia}
\email{streltsov@optics.imp.uran.ru}

\author{D.I.~Khomskii}
\affiliation{II. Physikalisches Institut, Universit$\ddot a$t zu K$\ddot o$ln,
Z$\ddot u$lpicher Stra$\ss$e 77, D-50937 K$\ddot o$ln, Germany}

\date{\today}

\begin{abstract}
The results of the LSDA+U calculations for pyroxenes with diverse magnetic properties (Li,Na)TM(Si,Ge)$_2$O$_6$,
where TM is the transition metal ion (Ti,V,Cr,Mn,Fe), are presented. 
We show that the anisotropic orbital ordering results in the spin-gap formation
in NaTiSi$_2$O$_6$. The detailed analysis of different contributions to the
intrachain exchange
interactions for pyroxenes is performed both analytically using perturbation
theory and basing on the results of the band structure calculations. 
The antiferromagnetic $t_{2g}-t_{2g}$ exchange is found to decrease gradually in
going from Ti to Fe. It turns out to be nearly compensated by
ferromagnetic interaction between half-filled $t_{2g}$ and empty $e_g$
orbitals in Cr-based pyroxenes. The fine-tuning of the interaction parameters
by the crystal structure results in the ferromagnetism for
NaCrGe$_2$O$_6$. Further increase of the total number of electrons
and occupation of $e_g$ sub-shell makes the
$t_{2g}-e_g$ contribution and total exchange interaction antiferromagnetic
for Mn- and Fe-based pyroxenes. Strong oxygen polarization
was found in Fe-based pyroxenes. It is shown that this effect leads
to a considerable reduction of antiferromagnetic intrachain exchange. The obtained 
results may serve as a basis for the analysis of diverse magnetic properties of pyroxenes, 
including those with recently discovered multiferroic behavior.   
\end{abstract}

\maketitle

\section{Introduction \label{intro}}
 Pyroxenes with the general formula AMX$_2$O$_6$ are an extremely rich class of 
compounds. Here $A$ may be alkali Na, Li or alkaline-earths Ca, Sr,... elements, $M–$-- 
different metals with the valence 3+, $X$ is most typically Si$^{4+}$, but also 
Ge$^{4+}$  and even could be V$^{5+}$.
These systems are best known in geology and mineralogy: they are one of the main 
rock-forming minerals in the Earth's crust, and are also found on the Moon, 
planets and in meteorites. 
One of the pyroxenes, NaCrSi$_2$O$_6$, may be  
found in natural form only in meteorites, which is reflected in its name: 
mineral kosmochlore. 
Another one, NaAlSi$_2$O$_6$, is the famous Chinese jade.
For the condensed matter physics the pyroxenes with $M$  a transition metal (TM), 
are of particular interest because of their rich and interesting magnetic 
properties.

\begin{center}
\begin{figure}[b!]
 \centering
\includegraphics[clip=false,width=0.4\textwidth]{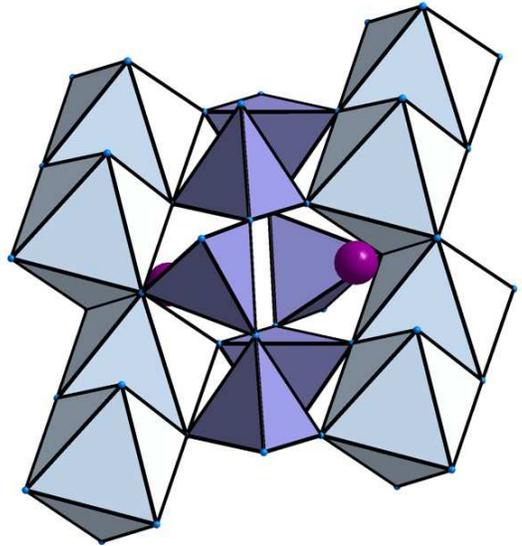}
\caption{\label{crystal-structure}(color online) Crystal structure of 
pyroxenes.
Two chains made of TMO$_6$ octahedra and its connections via $XO_4$ 
(where $X$ is usually Si or Ge) tetrahedra are shown. Purple balls 
represent Li(Na) ions.} 
\end{figure}
\end{center}

\begin{center}
\begin{figure}[t!]
 \centering
 \includegraphics[clip=false,width=0.48\textwidth]{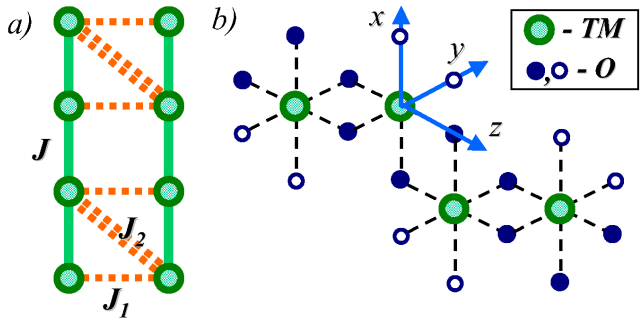}
\caption{\label{topology}(color online). (a) Triangular-like topology 
of the TM network in pyroxenes. Solid lines are intrachain links, 
dashed and double-dashed lines are the interchain connection via one or two 
XO$_4$ tetrahedra; $J$,
$J_1$ and $J_2$ are corresponding 
exchange parameters.
(b) Sketch of the one of TMO$_6$ chains. Two different types
oxygens are shown. Those depicted as solid circles belongs to two
different  TMO$_6$ octahedra, open circle - only to one. The
later oxygens are called ``side-oxygens'' in the text. The natural
choice of the local coordinate system (LCS), when $z-$axis is directed
along the single 180$^{\circ}$ O-TM-O bond connecting two neighboring  
TMO$_6$ octahedra, is shown.} 
\end{figure}
\end{center}

Structurally magnetic sub-system in pyroxenes is quasi-one-dimensional: TMO$_6$ 
octahedra form edge-sharing zig-zag chains connected by the chains of 
$XO_4$ tetrahedra (see Fig.~\ref{crystal-structure}). Two aspects are here especially important. First
of all the edge-sharing character with $TM-O-TM$ angle close, but not 
exactly equals to 90$^{\circ}$, which results in the competition between
different contributions to the super-exchange (see Sec.~\ref{exc-mechanism}). 
Secondly a ``shifted'' packing of neighboring
TM-chains leads to the situation, when every TM ion is connected 
(via own O and $XO_4$ tetrahedra) with two TM ions in neighboring chain. 
Thus, the general topology of exchange interaction is triangular-like, see 
Fig.~\ref{topology}(a), and the magnetic system may be frustrated. 

Both these factors may lead to nontrivial magnetic properties of pyroxenes, 
which are indeed observed experimentally, although for most of them the 
detailed studies are still absent. There are among them
antiferromagnetic (e.g. (Li,Na)V(Si,Ge)$_2$O$_6$~\cite{Vasiliev-04}, 
(Li,Na)FeSi$_2$O$_6$~\cite{Redhammer-01}) and even ferromagnetic 
(e.g. NaCrGe$_2$O$_6$~\cite{Vasiliev-05}) insulators. 
Some of the pyroxenes have the ground state with the spin gap: 
CaCuGe$_2$O$_6$~\cite{Sasago-52},  NaTiSi$_2$O$_6$~\cite{Isobe-02,Streltsov-06}, 
although the nature of such a state is apparently different: this seems to be 
due to formation of interchain dimers in CaCuGe$_2$O$_6$~\cite{Valenti-02}, 
and  due to the intrachain
dimerization in NaTiSi$_2$O$_6$, stabilized by the special type of orbital 
order~\cite{Isobe-02,Streltsov-06,Wezel-06}.

Finally a new twist in this story is the recent discovery 
that the pyroxenes apparently form a new class of multiferroics~\cite{Jodlauk-07}. 
It was found that
at least three of them, NaFeSi$_2$O$_6$, LiFeSi$_2$O$_6$ and LiCrSi$_2$O$_6$, develop an 
electric
polarization in a magnetically-ordered state, below $\sim 6$~K in 
NaFeSi$_2$O$_6$, $\sim 18$~K in LiFeSi$_2$O$_6$ and $\sim 11$~K in LiFeSi$_2$O$_6$. Similar to 
other 
multiferroics with magnetically-driven ferroelectricity (see e.g. 
reviews~\cite{Cheong-07,Khomskii-06}), electric polarization in them can be 
strongly modified by a magnetic field. Note also that the most pyroxenes are
insulating (often beautifully-colored crystals), which makes them especially
good for studying eventual ferroelectricity.

\noindent

To understand magnetic properties of this large and important class of compounds, 
and especially to have a better understanding of the possible mechanism of 
multiferroic behavior in them, 
it is desirable to know the details of 
the exchange coupling. The
knowledge of orbital occupancy is also very important, both
in connection with their structure and for determining the details of
exchange interactions. To reach this goal, and also to provide better understanding
of an electronic structure of pyroxenes in general, we undertook a detailed
theoretical investigation of these problems, concentrating on systems with 
$A$=Na, Li; $X$=Si,Ge and different trivalent TM ions. 
We carried out detailed ab-initio calculations, and in particular
obtained the values of the exchange constants. Besides these values themselves, 
which can be useful for interpreting the properties of real materials, we 
paid main attention to the study of the systematics of the exchange in different 
compounds of this large class, and to particular mechanisms, or partial
contributions of different exchange passes to the total exchange. 
General conclusions reached are important for the analysis of 
the multiferroic behavior in pyroxenes, they may be useful also for understanding 
of the properties of pyroxenes with alkaline-earths $A$ 
ions such as e.g. CaMnSi$_2$O$_6$ and may serve as a very good 
illustration of different tendencies in exchange couplings 
in systems with complicated geometry in general.

\section{Different contributions to exchange \label{exc-mechanism}}

Before presenting the results of our ab-initio calculations, let us first discuss 
different mechanisms of the intrachain exchange 
interaction. The explicit expressions for it may be calculated
using  the perturbation theory in 
$t/U_{dd}$ or $t/ \Delta$, where $t$ is an effective electron hopping 
(either direct $d-d$ hopping $t_{dd}$ or hopping between 
$d-$states of TM and $2p-$states of oxygen $t_{pd}$), $U_{dd}$ is
an on-site Coulomb interaction, and $\Delta$ is a charge-transfer 
energy (energy of the promotion of an electron from $2p-$states of 
oxygen to $3d-$states of TM).
\begin{center}
\begin{figure}[t!]
 \centering
 \includegraphics[clip=false,width=0.49\textwidth]{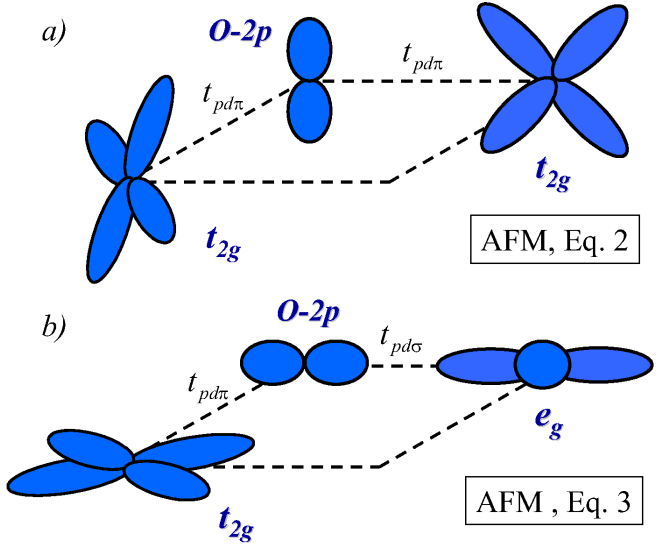}
\caption{\label{afm-oo}(color online). Typical orbital ordering
between two occupied TM $d-$orbitals via the same O$-2p$ orbital, which 
results to AFM interaction for the case of 90$^{\circ}$ $TM-O-TM$ bond.} 
\end{figure}
\end{center}
\noindent

The main motive of the intrachain packing of TM ions in pyroxenes is the 
edge-sharing packing of TMO$_6$ octahedra (see Fig.~\ref{crystal-structure}), with 
90$^{\circ}$ TM-O-TM bonds for (half)filled $t_{2g}- t_{2g}$ and $t_{2g}- e_{g}$ orbitals. 
This implies that one of the important contributions
will be a direct exchange between partially-filled $t_{2g}$ orbitals, 
the lobes of which look towards neighboring TM ions. This contribution to 
the total
exchange will be AFM and proportional to
\begin{eqnarray}
\label{J-direct}
 J_{t_{2g}-t_{2g}} \sim  \frac{t^2_{t_{2g}- t_{2g}}}{U_{dd}},
\end{eqnarray}
where $t_{t_{2g}- t_{2g}}$ is the $t_{2g}- t_{2g}$ hopping parameter
due to direct overlap of $d-$wave functions.
Since consecutive TM-O-TM planes in zig-zag TM-chains are rotated by 
90$^{\circ}$ (see Fig.~\ref{topology}b), specific orbital pattern will be crucial for this
contribution. This, in particular, leads to strongly alternating exchange in
NaTiSi$_2$O$_6$, determining the appearance of the spin gap in it 
(see Sec.~\ref{NaTiSi2O6}). On the other hand, corresponding orbital occupation
in V pyroxenes does not lead to an alternating exchange (Sec.~\ref{V}). And of course for
$t^3_{2g}$ occupation, as for Cr$^{3+}$  or Fe$^{3+}$, all the pairs 
between neighboring TM ions belonging to the same chain
would have equal contributions to this part of the exchange. 

Due to decrease of the radius of $d-$orbitals and with it of $t_{2g}-t_{2g}$ hopping,
this direct exchange is expected to strongly diminish in going from 
Ti to Fe pyroxenes, see below.
Note that we use the terminology ``direct exchange'' for the exchange
caused by the direct $d-d$ overlap and hopping, and reserve the 
term ``super-exchange'' for the exchange mediated by oxygens~\cite{direct-exch}. 

\begin{center}
\begin{figure}[t!]
 \centering
 \includegraphics[clip=false,width=0.49\textwidth]{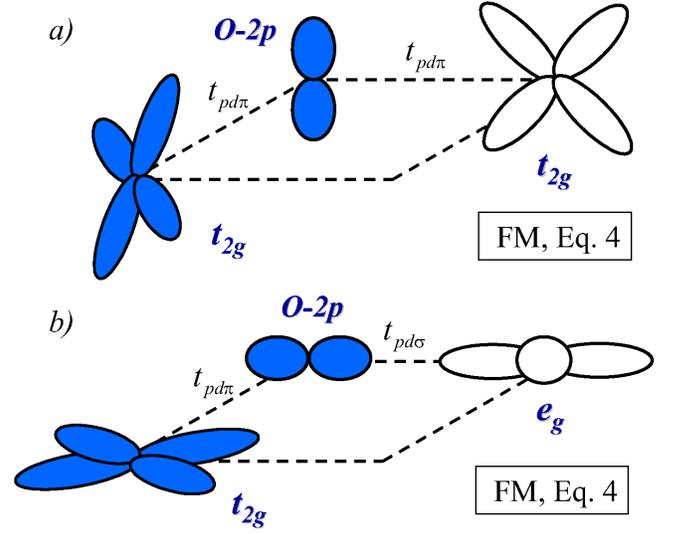}
\caption{\label{fm-empty-oo} (color online). Typical orbital ordering
between occupied (filled) and unoccupied (unfilled) $d-$orbitals 
of TM via the same O$-2p$ orbital, which results to FM
interaction for the case of 90$^{\circ}$ $TM-O-TM$ bond.} 
\end{figure}
\end{center}

\begin{center}
\begin{figure}[t!]
 \centering
 \includegraphics[clip=false,width=0.49\textwidth]{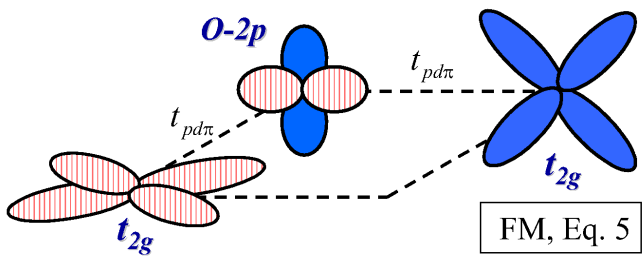}
\caption{\label{fm-dif-oo} (color online). Typical orbital ordering
between two occupied $d-$orbitals of TM via {\it different}
O$-2p$ orbitals, which results in a  FM
interaction for the case of 90$^{\circ}$ $TM-O-TM$ bond.} 
\end{figure}
\end{center}

Besides direct $d-d$ hopping, $p-d$ hybridization leads to an important
contribution to the total exchange due to TM-O-TM hopping. 
Different possible situations leading to a super-exchange via oxygens are 
illustrated in Figs.~\ref{afm-oo}-\ref{fm-dif-oo}.
 In these figures (half) occupied orbitals of TM are shown as 
filled (blue or hatched red), and empty orbitals - as empty (white) ones.
>From these figures it is clear that the corresponding contributions can be 
both antiferro- and ferromagnetic. 

The simplest case is that of Fig.~\ref{afm-oo}(a). 
Here half-filled orbitals at neighboring TM's overlap 
with the same oxygen $p-$orbital.
According to the usual rules~\cite{Goodenough} this 
antiferro-orbital ordering gives a substantial AFM exchange:
\begin{eqnarray}
\label{t2g-t2g-hf} 
J^{SE}_{t_{2g}-t_{2g}} \sim \frac{t_{pd \pi}^4}{\Delta^2}
\Big(\frac{1}{2\Delta+U_{pp}} + \frac{1}{U_{dd}}\Big); 
\end{eqnarray}
here $t_{pd \pi}$ is the $\pi-$hopping 
between $t_{2g}$ and $2p-$orbitals.
It is important that the magnitude
of $t_{2g}-O-t_{2g}$ ($e_g-O-e_g$) exchange presented in 
Eq.~(\ref{t2g-t2g-hf}) does not strongly depend on the 
TM-O-TM angle $\alpha$ (only via dependence of $t_{pd}$ hopping
on $\alpha$). First and second contributions are due to the correlation 
and delocalization effects in terminology of Ref.~\onlinecite{Goodenough}.

Similarly, in the case of Fig.~\ref{afm-oo}(b) half-filled 
$t_{2g}$ and $e_g$ orbitals 
overlap via the same oxygen orbital, which gives even stronger AFM exchange
\begin{eqnarray}
\label{t2g-t2g-hf2} 
J^{SE}_{t_{2g}-e_g} \sim \frac{t_{pd \pi}^2 t_{pd \sigma}^2}{\Delta^2}
\Big(\frac{1}{2\Delta+U_{pp}} + \frac{1}{U_{dd}}\Big), 
\end{eqnarray}
where $t_{pd\sigma}$ is the hopping between $e_g$ orbital of TM  and 
$2p-$orbital of oxygen directed towards it. $t_{pd\sigma}$ is approximately
two times larger than $t_{pd\pi}$~\cite{Harrison}.
\noindent

If the hopping is only possible from the occupied to an 
empty $d-$orbital, or if occupied orbitals of TM do not overlap 
with the same oxygen orbital, the resulting contribution to the 
exchange turns out to be ferromagnetic.

First of all let us consider the situation when one of $d-$orbitals is 
empty and it overlaps with the same oxygen $p-$orbitals, as the occupied
$d-$orbital, see Fig.~\ref{fm-empty-oo}(a,b). Then electrons can virtually
hop to this empty orbital, but the Hund's rule coupling $J_H^{TM}$
tends to orient spins at an ion  parallel. This will result in
a FM exchange: 
\begin{eqnarray}
\label{FM-empty}
J^{SE}_{t_{2g}-t_{2g}/e_g} \sim - \frac{t_{pdm}^2t_{pdm'}^2} {\Delta^2} 
\Big(\frac{J_H^{TM}}{(2\Delta +U_{pp})^2}
+ \frac{J_H^{TM}}{U_{dd}^2}\Big).
\end{eqnarray}
Note that $J_H^{TM}$ here is the Hund's energy gain 
when we put an extra electron with parallel spin to the original 
state of a TM ion, i.e. depending on the total spin of the latter 
$J_H^{TM}$ may be equal e.g. to twice the usual atomic value of 
$J_H \sim$0.8 eV for $3d-$series when the spin of the TM ion 
is $S=1$ (two electrons with parallel spins as in V$^{3+}$), or 
3$J_H$, as for Cr$^{3+}$ ($d^3$, $S=3/2$). This may enhance the 
ferromagnetic contribution in certain cases 
(in particular for Cr systems).

The last contribution to the total $d-d$ 
exchange interaction is coming from the overlap of
occupied $d-$orbitals with different 
(orthogonal) $2p-$orbitals. Typical orbital pattern for this situation
is shown in Fig.~\ref{fm-dif-oo}, but one may draw 
similar pictures also for the case of $e_g/e_g$ or $e_g/t_{2g}$
orbitals. This exchange will be FM and again because of the Hund's rule, but
in this case acting in oxygen $2p-$shell. 
With such an overlap the consecutive hopping of the electrons from
one $d-$ion to another one through oxygen $p-$shell
(delocalization exchange  
of Ref.~\onlinecite{Goodenough}) turns out to be
forbidden. Two oxygen's electrons are hopping to different
TM ions instead. Because of the Hund's coupling on oxygen ($J^p_H$)
the FM ordering becomes more favorable:  
\begin{eqnarray}
\label{diff} 
J^{SE}_{t_{2g}/e_g - t_{2g}/e_g } 
\sim - \frac{t_{pdm}^2t_{pdm'}^2J^p_H}  
{\Delta^2(2\Delta+U_{pp})^2}, 
\end{eqnarray}  
Note that $t_{pdm}$ in the expressions \eqref{FM-empty} and
\eqref{diff} may be $t_{pd\pi}$ or $t_{pd\sigma}$, depending on the particular
case. For instance, for the case depicted in 
Fig.~\ref{fm-empty-oo}(b) one them is $t_{pd\pi}$ and another
$t_{pd\sigma}$.

In order to describe magnetic interactions in a particular pyroxene
one has to include in general all possible contributions discussed above, 
via both common oxygens, and take into account several general trends.
\noindent

First of all, the radius of $d-$states ($r_d$) and with them the 
direct $d-d$ ($t_{dd}$) hoppings decrease going from the
light to heavy TM (from Ti, V to Fe).  According to the famous Harrison
scaling~\cite{Harrison},
\begin{eqnarray}
\label{Harrison-scaling}
t_{dd} \sim \frac{r_d^{3}}{D^{5}},
\end{eqnarray}
where $D$ is the distance between TM.
 Taking into account
typical values for $r_d$ presented in Ref.~\onlinecite{Harrison}
and assuming that $D$ is the same,
we found that $t^{Ti}_{dd} \sim 2.5 t^{Fe}_{dd}$. 
Thus, only this 
effect may provide suppression of the direct exchange in Fe-based
pyroxenes comparing with those based on Ti by a factor of 6.
Secondly, the distance between TM ions $D$ by itself strongly
affects the exchange coupling, as seen from Eq.~\eqref{Harrison-scaling}.
Third, there exist the strong increase of $U_{dd}$ and decrease of the charge-transfer
energy $\Delta$ for the late TM's. $\Delta$ enters the expressions 
Eq.~\eqref{t2g-t2g-hf}-\eqref{diff} in the denominators either as 
$\Delta^2$ or even as $\Delta^4$ and plays very important role.
For Ti$^{3+}$ $\Delta$ equals to 6~eV, whereas for Fe$^{3+}$ it is 
2.5~eV only~\cite{Bocquet-92}. 
This factor leads for the late 3$d-$elements to an increase of the
role of super-exchange via oxygen, which competes with
the direct $d-d$ exchange dominating in early TM's.

%Note, in addition that trasfering an 
%electron from oxygen to $e_g-$shell we spend not charge transfer energy
%$\Delta$, but also it needs to overcome $t_{2g}-e_{g}$ crystal-field
%splitting.

The qualitative discussion presented in this section will 
serve as a framework for interpreting the results of the
following sections. As we will see, all these factors are 
quite important and indeed determine the tendencies of exchange 
constants in pyroxenes and many other systems with 
90$^{\circ}$ $TM-O-TM$ bonds.

\section{Computational details \label{comp-retails}}
Crystallographic data used in the calculations were taken 
from the following papers: 
NaTiSi$_2$O$_6$ (T=100~K, space group P-1)~\cite{Redhammer-03},
LiVSi$_2$O$_6$ (T=293~K, space group C$2/c$)~\cite{Satto-97}, 
NaVSi$_2$O$_6$ (T=296~K, space group C$2/c$)~\cite{Ohashi-94}, 
LiCrSi$_2$O$_6$ (T=100~K, space group P$2_1/c$)~\cite{Redhammer-04}, 
NaCrSi$_2$O$_6$ (T=300~K, space group C$2/c$)~\cite{Origlieri-03}, 
NaCrGe$_2$O$_6$ (T=293~K, space group C$2/c$)~\cite{Redhammer-07},
NaMnSi$_2$O$_6$ (T=302~K, space group C$2/c$)~\cite{Ohashi-86},
NaFeSi$_2$O$_6$ (T=14~K, space group C$2/c$)~\cite{Ballet-89},
LiFeSi$_2$O$_6$ (T=100~K, space group P$2_1/c$)~\cite{Redhammer-01}.

There is a controversy in the literature about the crystal structure of
NaVGe$_2$O$_6$ and LiVGe$_2$O$_6$.
 Although all the studies were performed for
the room temperature, the difference in lattice parameters 
exceeds 0.3 $\AA$ for NaVGe$_2$O$_6$~\cite{Emirdag-04,Vasiliev-04}. 
There is also uncertainty about the space group for this 
material~\cite{Emirdag-04,Pedrini-04}. 
In contrast to NaVGe$_2$O$_6$, there is no striking
disagreement about the crystal structure of LiVGe$_2$O$_6$ in
Ref.~\onlinecite{Emirdag-04,Vasiliev-04}. However, we were not
able to reproduce distances presented in Ref.~\onlinecite{Emirdag-04}
from their crystal structure.  Therefore, the study of electronic and
magnetic properties of both NaVGe$_2$O$_6$ and LiVGe$_2$O$_6$ was
postponed for the future, when all the problems about 
the crystal structure of these compounds would be resolved.  

The primary calculations were performed within the framework of the 
linear muffin-tin orbitals method~\cite{Andersen-84}. 
The values of on-cite Coulomb interaction ($U$) and Hund's rule coupling
($J_H$) parameters were taken as following: $U_{Ti}=3.3$~eV, $U_{V}=3.5$~eV,
$U_{Cr}=3.7$~eV, $U_{Mn,Fe}=4.5$~eV; $J_H=0.8$~eV for Ti, V and
Cr, $J_H=0.9$~eV for Mn and $J_H=1$~eV for Fe.
The values of $U$ and $J_H$ for Ti were calculated for the same Ti radii
in Ref.~\cite{Streltsov-05}. For other TM those parameters were taken
as an average in wide variety of values of $U$ and $J_H$ for
TM$^{3+}$ ions presented in the literature. For the material
with small values of exchange parameters (cromates) we 
checked that the small variation of $U$ ($\pm 0.5 eV$) doesn't change 
magnetic ground state. 

Instead of LDA+U approximation used in Ref.~\onlinecite{Streltsov-06},
we utilized LSDA+U approach in the present paper. The difference between
these two methods is in the parts of the Hamiltonian
allowed to be spin polarized. In the LDA+U method only d-shell of the transition
metal ions may be spin-polarized; for all the other ions different spin-states 
are averaged out. This allows more easy account  
of the double counting in 
LDA+U formalism, but does not treat correctly the effects of possible spin polarization of oxygens, which significantly restricts the applicability of the method to more complex systems.
We found the noticeable spin polarization of oxygens for the late TM
pyroxenes (Mn, Fe, see Sec.~\ref{Mn}) 
in the LSDA+U calculations and used
this approach for the whole series. For the Ti and V systems the effects
of oxygen polarization is insignificant and these two approximations
give essentially the same results. For Fe-based pyroxenes LDA+U
overestimated exchange constants by as much as $\sim 25-30\%
$. 

The second difference from the Ref.~\onlinecite{Streltsov-06} is the more careful
choice of the MT-radii. We found that for not uniform atom distribution
in dimerized NaTiSi$_2$O$_6$ the LMTO calculation is very
sensitive to the radii of Na MT-sphere. As a result of too large $R_{Na}=3.5$~a.u. 
used in Ref.~\onlinecite{Streltsov-06}, Na ion takes the electrons in excess, which
leads to the significant overestimate of the exchange constant. In the present paper
we used $R_{Na}=3.3$~a.u. for all pyroxenes. This choice of $R_{Na}$ results in a
good agreement with the band structure and exchange constants obtained in the
pseudo-potential calculations, as it will be discussed in Sec.~\ref{NaTiSi2O6}.  
Note, however, that the main conclusions
about the importance of the strong Coulomb correlations made in Ref.~\onlinecite{Streltsov-06}
are valid, and do not depend on the choice of $R_{Na}$. In addition,
it is interesting enough that only the results for NaTiSi$_2$O$_6$ significantly
depend on the $R_{Na}$. Other pyroxenes do not show such a strong dependence.
For instance, by going from $R_{Na}=3.5$~a.u. to $R_{Na}=3.3$~a.u. for NaCrGe$_2$O$_6$,
the exchange constants change by 1.1~K, and for NaCrSi$_2$O$_6$ -- even less, 0.2~K. 
\begin{center}
\begin{figure}[t!]
 \centering
 \includegraphics[clip=false,width=0.4\textwidth,angle=270]{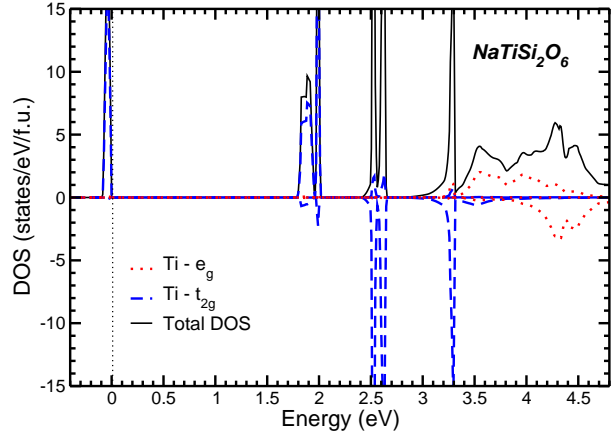}
\caption{\label{NaTiSi2O6-lda+u}(color online). 
Total and projected DOS for NaTiSi$_2$O$_6$ obtained
in LSDA+U calculation. Total DOS is the sum for both spins.
Up (down) panel corresponds to spin majority (minority).
Fermi level is zero energy.} 
\end{figure}
\end{center}
\noindent

The radii of other elements were set as following: $R_O=1.65-1.75$~a.u.,
$R_{Si,Ge}=1.9$~a.u., $R_{Li}=2.9$~a.u. The radii of transition metals were
chosen to be $R_{TM}=2.4-2.67$~a.u.

The Ti,V,Cr,Mn,Fe($4s$,$4p$,$3d$), O($2s$,$2p$,$3d$), Si($3s$,$3p$,$3d$), 
Ge($4s$,$4p$,$4d$), Li($2s$,$2p$,$3d$) and Na($3s$,$3p$,$3d$)
orbitals were included to the orbital basis set in our calculations. 
The Brillouin-zone (BZ) integration in the course of the
self-consistency iterations was performed over a mesh of 
64 {\bf
k}-points in the irreducible part of the BZ. For those
systems for which  exchange parameters are particularly small,
the calculation was checked by the finer mesh of 216 {\bf
k}-points.

In our notations, Heisenberg Hamiltonian is written in the following
form: 
\begin{equation}
\label{Heisenberg}
H = \sum_{ij}J_{ij} \vec{S_i} \vec{S_j},
\end{equation}
where summation runs twice over every pair $i,j$. $J$ were computed 
in the framework of Lichtenstein's exchange interaction parameter (LEIP)
calculation procedure~\cite{Katsnelson-00}.
According to the LEIP formalism, exchange constants $J$ can be 
calculated as the second derivatives of the energy variation at small
spin rotation. For $s=1/2$ 
\begin{eqnarray}
\label{LEIP}
J_{ij} = \frac{{\rm Im}}{\pi} \int \limits_{-\infty}^{E_{F}} d\epsilon 
\sum_{\substack {m m' \\ m'' m'''}}
\Delta^{mm'}_{i} \,
G_{ij \, \downarrow}^{m'm''} \, 
\Delta^{m'' m'''}_{j} \, G_{ji \, \uparrow}^{m'''m}, 
\end{eqnarray}
where $m$, $m'$, $m''$, $m'''$ are the magnetic quantum numbers, $i,j$ $-$ the lattice
indexes, $\Delta^{mm'}_{i}=H^{m m'}_{ii \, \uparrow} - H^{m m'}_{ii \, \downarrow}$ is the
on-site potential correction due to different spins
and the Green's function is calculated in the following way:
\begin{eqnarray}
\label{Green}
G^{mm'}_{ij \sigma}(\epsilon) \, = \, \sum_{k,\, n} \frac{c^{mn}_{i \sigma} \, (k) \, c^{m'n \, *}_{j \sigma} \,
(k)}{\epsilon-E^{n}_{\sigma}}.
\end{eqnarray}
Here $\sigma$ is the spin index, $c^{mn}_{i\sigma}$ is a component of the 
{\it n}-th eigenstate, and E$_{\sigma}^{n}$ is the corresponding eigenvalue.
Since transition metal ions are usually magnetic, the Green
functions $G^{mm'}_{ij \sigma}(\epsilon)$ and potential corrections $\Delta^{mm'}_{i}$
are computed only for them.

The great advantage of the LEIP procedure is the possibility to calculate not only 
total $J$'s between corresponding sublattices of transition metal ions, but, 
since it is linear with respect to the magnetic quantum numbers, 
one can also compute partial contributions, which come from
different $d-$orbitals. 

These orbitals were defined in the local coordination system (LCS), where axes
are orthogonal and directed as much as possible to the neighboring oxygens. 
The natural choice of the LCS is when the local $z-$axis is directed
along the only 180$^{\circ}$ $O-TM-O$ connecting two neighboring
TMO$_6$ octahedra, see Fig.~\ref{topology}(b). This definition
is unique, and it is used for all the pyroxenes studied in the present paper.

Since some of the exchange constants obtained within the LMTO method using LEIP
were found to be sensitive to the choice of MT-radii, in the present
paper we rechecked the LMTO results by calculating the exchange constants from 
the total energies in the pseudo-potential PW-SCF code~\cite{PW-SCF}. 
The formalism of pseudo-potential method does not require the introduction of
any atomic spheres. We used ultrasoft pseudo-potentials with nonlinear core
correction taken from www.pwscf.org with different forms of exchange-correlation.
The plane-wave and kinetic energy cut-off's were chosen to be 40 Ry and 
200 Ry, respectively. 216 $k-$points were used in the course of self-consistency.

Hopping integrals were obtained using Wannier function projection
in LMTO method described elsewhere~\cite{Anisimov-05, Streltsov-05}. Small subspace
defined by three $t_{2g}$ orbitals (defined in LCS) per V atom was used
in this procedure. Since for all of the considered pyroxenes $t_{2g}$-bands
are well isolated, resulting projected and initial LDA bands completely
coincide.

\section{\label{NaTiSi2O6} T\lowercase{i}-based pyroxene: 
N\lowercase{a}T\lowercase{i}S\lowercase{i}$_2$O$_6$ ({\it \lowercase{d}$^1$})}

As it was discussed in details in Sec.~\ref{intro}, pyroxenes are made 
of isolated zig-zag chains of TMO$_6$ octahedra. Combination of low
dimensionality together with the small value of the spin moment 
in NaTiSi$_2$O$_6$ results in an opening of the spin-gap
~\cite{Isobe-02,Baker-07}. Here we will show that it
becomes possible due to a particular orbital ordering.    
\begin{center}
\begin{figure}
 \centering
 \includegraphics[clip=false,width=0.5\textwidth]{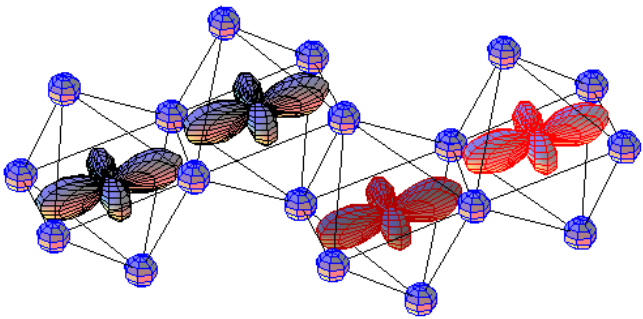}
\caption{\label{OO-NaTiSi2O6} (color online). The orbital ordering
obtained in LSDA+U calculations for NaTiSi$_2$O$_6$ is shown.} 
\end{figure}
\end{center}
\noindent

Two different models were suggested 
in order to explain the physical origin of the
spin-gap formation. The first one (dimer model) was initially proposed in
experimental papers~\cite{Isobe-02,Konstantinovich-04} 
and supported
afterward by the LDA+U calculations~\cite{Streltsov-06}.
It takes into account quite strong dimerization of Ti-Ti chain found 
in very detailed structural study performed by Redhammer and co-authors 
~\cite{Redhammer-03}. According to this model 
the ground state of NaTiSi$_2$O$_6$ consists of  spin singlet Ti-Ti dimers, and the 
spin-gap observed in the experiment is the energy gap between 
spin singlet and triplet states of these dimers.

In another model (Haldane scenario)~\cite{Popovich-04, Popovich-06} Ti-Ti dimers 
were proposed to be in a
spin triplet state and such $S=1$ Ti-Ti pairs form the AFM
chain, which is known as Haldane chain~\cite{Haldane-83} with the
spin-gap $\Delta_{SG} = 0.41 J$, where $J$ is the exchange
constant. 

Thus, the key point in these two scenarios is the degeneracy of the magnetic ground
state of Ti-Ti dimer: in the dimers model it is a singlet, but in the
Haldane one - a triplet.

It is interesting to note that depending on the situation the pair of  
metal ions may have quite different degeneracy of the magnetic ground state. In 
the case of isolated cluster of metal atoms in 
vacuum, the ions in the pair are very tightly bound and must be considered as 
one quantum-mechanical
object. The bond length of such isolated ``metal-metal molecule''
respectively is quite small: 1.9~$\AA$~for Ti-Ti pair~\cite{Gutsev-03}.
The spin configuration in this case will be governed by the Hund's coupling,
which favors the state with larger multiplicity. As a result
the ground state of isolated metal-metal dimers usually is {\it not}
the singlet one~\cite{Gutsev-03}.
 
However the situation in insulating transition metal oxides is qualitatively different.
Typical metal-metal distance varies here depending on the type
of the crystal structure, but usually it is much larger than
in isolated dimers: Ti-Ti distance is 3.1~$\AA$  in
the edge-sharing octahedral structure of NaTiSi$_2$O$_6$~\cite{Redhammer-03}.
As a result every ion has its own ground state configuration, and the influence
of the other ions can be incorporated via perturbation 
theory~\cite{Goodenough}. The resulting
electronic Hamiltonian can be reduced to a Heisenberg model~\cite{Kugel-82},
which is known to have spin-singlet ground state for the dimer.
Our band-structure calculations  in LSDA+U approximation give the same result:
rather strong antiferromagnetic exchange in short Ti-Ti dimers $J_{intra}=396$~K,
and the weak exchange between dimers $J_{inter}=5$~K.

The electronic structure of NaTiSi$_2$O$_6$, obtained in the LSDA+U approach is
presented in Fig.~\ref{NaTiSi2O6-lda+u}. Both the top of the valence band
and bottom of conduction one are formed by Ti-$3d$ states,
divided by the band gap of 1.77~eV. Singly occupied Ti-$3d$ orbital
has $zy-$character. Corresponding ferro-orbital ordering
is shown in Fig.~\ref{OO-NaTiSi2O6}. In a short Ti-Ti pair it results 
in a strong direct exchange interaction, which is AFM in agreement with 
Goodenough-Kanamori-Anderson (GKA) rules
~\cite{Goodenough,Kugel-82}. On the other hand, for the long Ti-Ti pairs the
orbitals are directed in such a way that the overlap between them
is almost zero, which leads to a rather small inter-dimer exchange.

Note that the exchange constants presented above are different from
those given in Ref.~\onlinecite{Streltsov-06}. This is due to a more careful
choice of Na MT-radii, as it was discussed in Sec.~\ref{comp-retails}.
In the present paper we additionally checked the LMTO results by calculating
total energies using pseudo-potential PW-SCF code~\cite{PW-SCF}.  In order to
have reliable result the calculations with the use of two different
pseudo-potentials based on Perdew-Wang-91 (PW91) and Perdew-Bruke-Ernzerhof(PBE)
exchange potentials were utilized. The total energy difference between
FM and AFM solutions were found to be 398~K for PW91 and 395~K
for PBE, which are very closed to the present LMTO results.      

One should be careful comparing exchange constants obtained in the band structure
calculation with the value of the spin gap.
This is due to strong quantum effects. What is indeed calculated by
any mean-field method are the total energies for AFM and FM configurations
(or the change of these energies with respect to the small
rotation angle of spins, as in LEIP, which is essentially the same).
For the case of isolated dimer $s=1/2$ and Heisenberg model defined by
Eq.~\eqref{Heisenberg} the total energy difference between FM and AFM
configuration equals to $J$. The spin gap $\Delta_{SG}$ is the energy difference between
{\it quantum} singlet and triplet states, which are
not the same as collinear AFM and FM states. For the $s=1/2$ the spin-gap 
$\Delta_{SG}=2J$.
The spin-gap obtained for NaTiSi$_2$O$_6$ using different experimental
methods is about 600-700~K~\cite{Baker-07}. This agrees 
with $\Delta_{SG} \sim 790$~K calculated in the presented paper.

\section{V-based pyroxenes ({\it\lowercase{d}$^2$}) \label{V}}
Ionic configuration of V in the pyroxenes  is $d^2$ implying 
$S=1$. Because of a larger spin moment it is much harder to make dimerized chains
in V-based pyroxenes. The spin gap may still appear in the
system, if it could be considered as a set of isolated one-dimensional
AFM chains of $S=1$ (Haldane chains~\cite{Haldane-83}).
However, experimentally all the V-pyroxenes show 
long-range magnetic ordering at low temperatures~\cite{Vasiliev-04}. 
Three different reasons for it were discussed in the literature:
biquadratic exchange~\cite{Millet-99,Gavilano-00}, interchain exchange
~\cite{Lumsden-00, Pedrini-07} and 
next-nearest-neighbor coupling~\cite{Gavilano-00}.

Common structural feature for all of the pyroxenes except Mn-based ones
(they are Jahn-Teller active) is the presence of two short TM-O bonds. 
These are the bonds with ``side-oxygens'', see Fig.~\ref{topoloy}(b).

Short V-O bonds form nearly 90$^{\circ}$ angle. Such a distortion of
TMO$_6$ octahedra
may be represented as two independent compressions in $x$ and $y$
directions. As a result $zx-$ and $zy-$orbitals will be almost 
degenerate and have lower energy than $xy$.

\begin{table}
\centering \caption{\label{J-table}
Results of the LSDA+U calculations for different pyroxenes.
The values of the band gap and spin moments were taken
from AFM (intrachain) solution intrachain calculations
for all the compounds, except NaCrGe$_2$O$_6$. The
exchange parameters were obtained in LEIP formalism} 
\vspace{0.2cm} 
\begin{tabular}{lcccccc}
\hline
\hline
 & Ionic conf.& Band     & Spin             & Exchange        \\
 & of TM      & gap (eV) & mom. ($\mu_B$) & intrachain (K)  \\ 
 \hline
NaTiSi$_2$O$_6$ & $d^1$ & 1.77 & 0.92  & 396   & \\
\hline
LiVSi$_2$O$_6$  & $d^2$ & 1.92 & 1.87  & 50.3 &   &\\
NaVSi$_2$O$_6$  & $d^2$ & 1.80 & 1.85  & 15.0 &   &\\
\hline 
LiCrSi$_2$O$_6$ & $d^3$ & 3.57 & 2.78 &  3.4 \\
NaCrSi$_2$O$_6$ & $d^3$ & 3.03 & 2.83 & -0.8 \\
NaCrGe$_2$O$_6$ & $d^3$ & 2.88 & 2.84 & -5.2 \\
\hline
NaMnSi$_2$O$_6$ & $d^4$ & 1.76 & 3.69   & 3.4 \\
\hline
LiFeSi$_2$O$_6$  & $d^5$ & 1.63 & 4.02  &
15.8\footnote{Note that because of the large O$-2p$
polarization the exchange parameters obtained in
LEIP formalism are significantly overestimated
for Fe-based pyroxenes; see Sec.~\ref{iron}.} \\
NaFeSi$_2$O$_6$  & $d^5$ & 2.31 & 4.24  & 15.9$^a$  \\
\hline
\end{tabular}
\end{table}
 
In our LSDA+U calculations two 3$d$ electrons of V follow the crystal-field
splitting and occupy $zx-$ and $zy-$orbitals,
i.e. these systems do not have orbital degeneracy.
 Each orbital provides strong
AFM direct exchange with one of the neighboring V. This leads, as
we will show below, to the uniform exchange interaction in each chain, with
the direct exchange defined in Eq.~\eqref{J-direct}. The super-exchange via
oxygens would be also homogeneous and (weaker) FM.

In our calculations all the (Li,Na)VSi$_2$O$_6$ pyroxenes were found 
to be insulators with the band gap $1.8 - 1.9$~eV. 
Spin moments are slightly
reduced from ionic value of 2$\mu_B$ because of the hybridization. 
Interestingly, the calculated intrachain exchange constants significantly
decrease going from  
Li~$\mapsto$~Na (see Tab.~\ref{J-table}).
In order to understand the reasons for such tendency, we took advantage of
the LEIP and calculated partial contributions to the total exchange constants.

We found that as in NaTiSi$_2$O$_2$, in V-pyroxenes the most important
contribution is the direct exchange interaction between $t_{2g}$ orbitals.
For LiVSi$_2$O$_2$ $J_{t_{2g}-t_{2g}}=59.6$~K. The
$t_{2g}-e_{g}$ partial exchange is FM, since
it occurs from occupied to empty orbital. 
Numerically this contribution is defined by Eq.~\eqref{FM-empty},
and in our LSDA+U calculation for LiVSi$_2$O$_2$
$J_{t_{2g}-e_{g}}=-9.9$~K. The rest, $0.6$~K (total
exchange for LiVSi$_2$O$_2$ is 50.3~K) is the partial exchange
between $e_g$ orbitals, which should be completely empty
in the purely ionic picture, but which still are partially occupied in the real
band structure calculation because of the hybridization with oxygens.

Going from Li to Na, the $t_{2g}-t_{2g}$ partial exchange weakens
because of the increase of the distances and local distortions of the
octahedra. For NaVSi$_2$O$_2$ $J_{t_{2g}-t_{2g}}=27.5$~K, 
$J_{t_{2g}-e_{g}}=-13.3$~K and $J_{e_{g}-e_{g}}=0.8$~K.
The direct calculation of hopping 
integrals corroborates this interpretation. Thus e.g. for LiVSi$_2$O$_6$
and NaVSi$_2$O$_6$ the ratio of $t_{2g}-t_{2g}$ exchange integrals
$J^{Li}_{t_{2g}} / J^{Na}_{t_{2g}} = 2.17$, which is close to the squared ratio
between effective $t_{2g}-t_{2g}$ hopping integrals $(t^{Li}/ t^{Na})^2 =2.15$.

The total values of intrachain interaction obtained within LMTO method 
$J=50.3$~K (LiVSi$_2$O$_6$) and $J=15.0$~K (NaVSi$_2$O$_6$) agree
with those calculated in pseudo-potential code from the total energies:
$J=51.3$~K (LiVSi$_2$O$_6$) and $J=13.8$~K (NaVSi$_2$O$_6$).
In addition it matches the exchange constants obtained
in Ref.~\onlinecite{Pedrini-07}, where the fitting of experimental curves 
of magnetic susceptibility to theoretical one, obtained using Quantum 
Monte Carlo simulations on a cubic lattice, was performed. Note however,
that Pedrini {\it et al}.\cite{Pedrini-07} used different definition
of Heisenberg Hamiltonian, and their exchanges must be divided by factor two
to compare with ours. In addition it should be mentioned that the real
geometry of pyroxenes includes frustrating diagonal exchange
paths depicted in Fig.~\ref{topology}(a) as dashed lines. Those contributions
were not taken into account in Ref.~\onlinecite{Pedrini-07}.

\section{C\lowercase{r}-based pyroxenes ({\it \lowercase{d}$^3$})}
\begin{center}
\begin{figure}[t!]
 \centering
 \includegraphics[clip=false,width=0.4\textwidth]{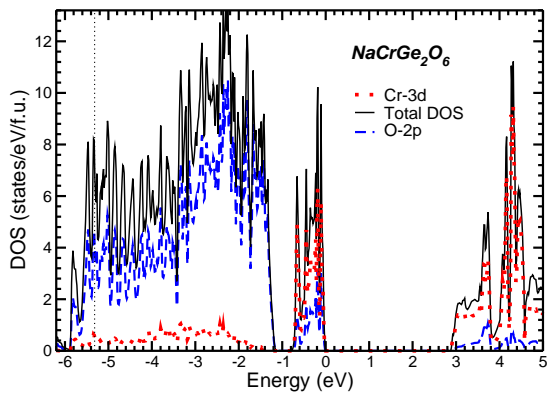}
\caption{\label{NaGrGe2O6-lda+u}(color online). 
Total and projected DOS for NaCrGe$_2$O$_6$ obtained
in LDA+U calculation. Fermi level is zero energy.}
\end{figure}
\end{center}
\noindent

Generally the electronic properties of Cr-based 
($d^3$ configuration, $S=3/2$) pyroxenes are similar to 
those of other pyroxenes: they are also insulators, but
with larger band gap $\sim$3~eV. This is mainly
because of the fact that spin-majority $t_{2g}$ sub-shell
becomes fully occupied and goes lower in energy. More important
is that the character of the valence band in Cr-based pyroxenes
is different.

 Both the Ti- and V- based pyroxenes were found to be Mott-Hubbard
insulators, with the gap separating occupied and unoccupied
transition metal 3$d$ bands. Going from the left to the
right in the same row of the periodic table, the charge-transfer
energy decreases, O-2$p$ states shift up~\cite{Bocquet-92}
and get to the same energy region as the lower transition metal
Hubbard band. This is clearly seen from the DOS plot obtained for NaCrSi$_2$O$_6$
in LSDA+U calculation and presented in Fig.~\ref{NaGrGe2O6-lda+u}.
Quite similar situation occurs in other Cr-based transition
metal oxides: one may observe significant mixing between O$-2p$ and
Cr$-3d$ states at the top of the valence band~\cite{Korotin-98,Streltsov-07}.
This feature of the electronic structure is important
for the magnetism of Cr-based pyroxenes, since it 
increases super-exchange interaction via O$-2p$ states. 

Experimentally LiCrSi$_2$O$_6$ and NaCrSi$_2$O$_6$ were found
to be AFM with small Neel temperature, 11~K and 3~K
correspondingly; NaCrGe$_2$O$_6$ is FM with $T_C=6$~K. 

In our calculations LiCrSi$_2$O$_6$ was indeed found
to be AFM with $J=3.4$~K. NaCrSi$_2$O$_6$ shows rather small
intrachain exchange, less than 1~K. This number is obviously lying
beyond the precision of the calculation
scheme~\cite{precision}. One may just note that 
this compound is on the borderline between FM and AFM ordering.
In agreement with the experiment,
intrachain exchange in NaCrGe$_2$O$_6$ is FM and more pronounced,
$J=-$5.2~K ($-7.9$~K in pseudo-potential calculation). 

 More detailed investigation
of the partial contributions to intrachain exchange shows that 
$t_{2g}-e_{g}$ exchange interaction in all Cr-based pyroxenes
is almost the same as in V-based ones: $J^{SE}_{t_{2g}-e_g} \sim$ -10.5~K. 
The AFM $t_{2g}-t_{2g}$ contribution, in contrast, is significantly
reduced in Cr pyroxenes. This seems to be the joint effect of the   
increase of the TM-TM distances and on-cite Coulomb repulsion $U$ and the
decrease of the charge-transfer energy $\Delta$ and radii of localization
of $d-$electron $r_d$, as it was discussed in the end of
Sec.~\ref{exc-mechanism}.

The largest $t_{2g}-t_{2g}$ exchange was found in LiCrSi$_2$O$_6$:
$J_{t_{2g}-t_{2g}}=$13.4~K, which has shortest $Cr-Cr$ distance 
($D=3.06$~$\AA$). For NaCrSi$_2$O$_6$ the $t_{2g}-t_{2g}$ exchange
is smaller, $J_{t_{2g}-t_{2g}}=$7.7~K, which is presumably related
to longer $Cr-Cr$ distance ($D=3.086$~$\AA$). Thus, according to
our calculations the FM $t_{2g}-e_g$ exchange almost compensates
AFM $t_{2g}-t_{2g}$ contribution in NaCrSi$_2$O$_6$.
This tendency continues for NaCrGe$_2$O$_6$, where because
of larger Ge and Na ionic radii the $Cr-Cr$ distance exceeds its
maximum value ($D=3.142$~$\AA$), and the FM $t_{2g}-e_g$ contribution
becomes dominating, since $J_{t_{2g}-t_{2g}}=3.1$~K.

\subsection{Changes of the electronic structure going from Si to
Ge on the example of Cr-based pyroxenes}
The Cr$-$based pyroxenes were the only pyroxenes with available and
reliable crystal structure to study the effect of $Si \leftrightarrow Ge$ 
interchange in these compounds. We studied only these, but argue
that the general character of the changes in the electronic structure
should be the same for other pyroxenes. 

There are two possible mechanism, which may determine
the difference in the electronic properties of Ge and Si based
pyroxenes. First of all Ge$-4p$ states have larger
spatial extensions than Si$-3p$, this may result in larger
covalency effects. Secondly the size of Ge$^{4+}$ ions is much larger 
than Si$^{4+}$ (the ionic radius $R^{IV}_{Ge^{4+}} = 0.39$~$\AA$, while 
$R^{IV}_{Si^{4+}} = 0.24$~$\AA$). Therefore, different structural
distortions are expected for Ge and Si pyroxenes.
\begin{center}
\begin{figure}[t!]
 \centering
 \includegraphics[clip=false,width=0.4\textwidth,angle=270]{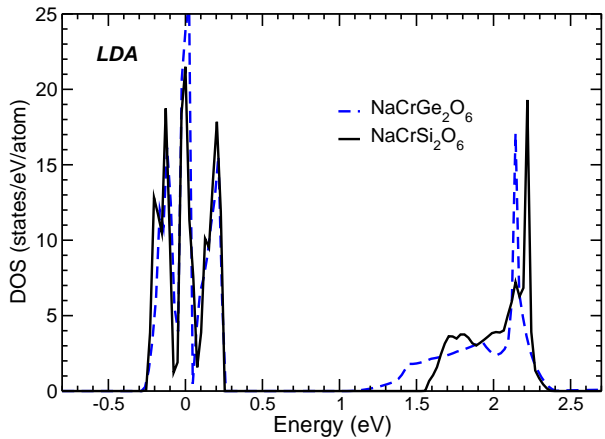}
\caption{\label{SiGe}(Color online). 
Cr$-3d$ DOS for NaCrSi$_2$O$_6$ and NaCrGe$_2$O$_6$
obtained in LDA. It illustrates the decrease of the $t_{2g}-e_{g}$
crystal-field splitting with substitution Si by Ge.
Fermi level is zero energy.} 
\end{figure}
\end{center}
\noindent

In order to understand how important is the effect of $4p$ or $3p$
states, we performed the calculations, where in the  NaCrGe$_2$O$_6$
Ge ions were substituted by Si. In this case all the distances
in ``artificial NaCrSi$_2$O$_6$'' were taken the same as in the parent
NaCrGe$_2$O$_6$. All the radii of MT-spheres were also taken the
same as for NaCrGe$_2$O$_6$. The results show minor
difference in shape and positions of Cr$-d$ bands compared to NaCrGe$_2$O$_6$. 
Therefore one may conclude that the origin
of the difference in the electronic properties between Si and Ge
pyroxenes is mostly due to different local distortions.   

The average $Ge-O$ bond distance ($d_{Ge-O}=1.74$~$\AA$ in NaCrGe$_2$O$_6$ ) 
is larger than $Si-O$ one ($d_{Si-O}=1.63$~$\AA$ in NaCrSi$_2$O$_6$ ).
Since the TM-O distance is mostly determined by the strong ionic TM-O bond
and hence by  the TM and O ionic radii only, the substitution
of Si by larger Ge leads to the bending of TM-O-TM bond.
If in NaCrSi$_2$O$_6$ the angle $Cr-O-Cr=99.6^{\circ}$, 
then in NaCrGe$_2$O$_6$ this angle is 
$101.2^{\circ}$. This 
leads to an increase of the Cr$-e_g$ bandwidth, while
the Cr$-t_{2g}$ bandwidth is not changing, as one may see in
Fig.~\ref{SiGe}. 

On the other hand, shorter $Si-O$ bond distance results
in the larger shift of the O$-2p$ states to the lower
energies, than in the case Ge. This should lead to an increase of the
crystal-field splitting in TM$-3d$ shell going from
Ge to Si.
The direct center of gravity calculation shows that $t_{2g}-e_g$ 
crystal-field splitting in NaCrSi$_2$O$_6$ equals to 1.89~eV, while
for NaCrGe$_2$O$_6$ it is 1.68~eV.

\section{M\lowercase{n}-based pyroxenes ({\it \lowercase{d}$^4$}) \label{Mn}}
We were able to find the crystal structure for only one Mn-based
pyroxene: NaMnSi$_2$O$_6$ (mineral namansilite). Having one electron in $e_g$ sub-shell,
Mn ion is Jahn-Teller (JT)-active. As a result the ``distribution''
of long and short Mn-O bonds in MnO$_6$ octahedron is different from
other pyroxenes. 

It is known that because of the anharmonic effects the JT distortion
around Mn$^{3+}$ ions typically leads to a local elongation of MnO$_6$ octahedra 
~\cite{Khomskii-00}. These elongated octahedra may in principle be packed
differently, leading both to an antiferrodistortive (for example as in LaMnO$_3$) and to a 
ferrodistortive (e.g. Mn$_3$O$_4$) ordering. 

In the edge-sharing geometry the largest gain in the
elastic energy may be achieved, if the longest Me-O bonds in neighboring
octahedra are parallel to each other. The explicit expressions for the
elastic energies of different distortion modes are summarized
in Tab.~1 of Ref.~\onlinecite{Khomskii-03}. Among various possible types
of parallel JT distortions in the edge-sharing octahedra, the
consecutively ``twisting'' geometry of pyroxenes selects the only
variant: the distortion occurs along the O-Mn-O bond connecting
two neighboring octahedra (local $z-$ axis). Only in this case the systems
gains the energy for both neighbors. The local crystal-field due to the
presence of ``side-oxygens'' forming short Mn-O bond, directed
only to the SiO$_4$ tetrahedra, stabilizes single $e_g$ electron on the
$3z^2-r^2$ orbital and therefore works in the same directions.

This conclusion is in  full agreement with the crystal-structure
studies performed in Ref.~\onlinecite{Ohashi-86}, where 
the parallel 180$^0$ O-Mn-O bonds along the local $z-$axis
connecting neighboring MnO$_6$ octahedra were found to
be the longest.

In the LSDA+U calculations we indeed obtained that on
all of Mn$^{3+}$ ions the single $e_g$ electron localizes on the
same $3z^2-r^2$ orbital, directed along this longest Mn-O bond.
Resulting magnetic moment equals to 3.69~$\mu_B$, the bang gap 
$\sim 1.76$~eV. The intrachain exchange calculated in LEIP procedure is 
AFM and $\sim 3.4$~K (from the total energy
difference in pseudo-potential code we obtained that $J=-0.1$~K)

The $t_{2g}-t_{2g}$ and $t_{2g}-e_g$ contributions to the total exchange
are AFM and equals to 2.1~K and 6.1~K respectively. 
The situation with half-filled $t_{2g}-t_{2g}$ sub-shells
is the same as in the case of Cr, but $t_{2g}-e_g$ exchange will now
predominantly go from the {\it occupied} $t_{2g}$ to {\it occupied} 
$e_g$ orbitals, as shown in Fig.~\ref{afm-oo}(b). 
This contribution should be AFM, as follows from Eq.\eqref{t2g-t2g-hf2}.
The $e_g - e_g$ contribution is about $-4.8$~K FM.

\section{F\lowercase{e}-based pyroxenes ({\it \lowercase{d}$^5$}) \label{iron}}
The Fe$^{3+}$ ions have the simple configuration $d^5(t_{2g}^3e_g^2)$, 
without any orbital degeneracy, thus the exchange is uniform
along the chain.
Using LEIP formalism we obtained that the intrachain exchange is AFM both in
LiFeSi$_2$O$_6$ ($J$=15.8~K) and NaFeSi$_2$O$_6$ ($J$=15.9~K). This is consisted
with neutron measurement performed in Ref.~\onlinecite{Redhammer-01} for LiFeSi$_2$O$_6$,
but disagrees with the experimental findings for 
NaFeSi$_2$O$_6$ presented in Ref.~\onlinecite{Ballet-89},
where Fe-ions were claimed to be ferromagnetically ordered
in the chain.

The $t_{2g}-t_{2g}$ partial contribution to the exchange in Fe pyroxenes
was found to be AFM and of order of that for Cr-based ones:
$0.5$~K and $1.4$~K for LiFeSi$_2$O$_6$ and NaFeSi$_2$O$_6$
respectively. The $t_{2g}-e_g$ exchange is also AFM since
it goes from the {\it occupied} $t_{2g}$
to {\it occupied} $e_g$ orbitals, as in the case of
NaMnSi$_2$O$_6$. It equals to $9.4$~K and $7.0$~K, respectively.

The situation with $e_{g}-e_g$ contribution here is
a bit different from the case of NaMnSi$_2$O$_6$.
The TM-O-TM angle equals to 
{$\sim 97.1^{\circ}$ in NaMnSi$_2$O$_6$, while in NaFeSi$_2$O$_6$ 
and LiFeSi$_2$O$_6$ it is larger: $\sim 100.5^{\circ}$ and 
$\sim 99.0^{\circ}$ respectively. This may be significant, since in deviating
from 90$^{\circ}$, small FM contribution given by Eq.~\eqref{diff}
is quickly outbalanced by much stronger AFM exchange of order $t^2/U_{dd}$.
Detailed analysis shows that the borderline, where the mutual
compensation occurs, is about 97$^{\circ}$~\cite{Geertsma-96}.
Thus, one may expect that in Fe-based pyroxenes $e_{g}-e_{g}$
exchange will be already AFM. Note however that such a strong
angle dependence is not a common feature of all the 90$^{\circ}$ 
contributions described in Sec.~\ref{exc-mechanism}. For instance,
the important for the cromates $t_{2g}-e_{g}$ exchange is almost 
angle independent.

Our LEIP calculation confirms this qualitative discussion.
The $e_{g}-e_{g}$  contribution in $Fe$ pyroxenes is indeed AFM and equals to 
$5.9$~K and $7.4$~K for LiFeSi$_2$O$_6$ and NaFeSi$_2$O$_6$
respectively.

Thus, from a theoretical point of view it is hard to expect 
that the intrachain exchange would be FM, as the
authors of Ref.~\onlinecite{Ballet-89} proposed.  
We see that the exchange constants and their partial contributions
obtained in LEIP formalism agree with the theoretical conclusions
based on the perturbation theory. However, they are obviously too large to
describe experimentally observed paramagnetic Curie temperatures:
$\Theta = -25.8$~K for LiFeSi$_2$O$_6$~\cite{Redhammer-01} and
$\Theta = -46$~K for NaFeSi$_2$O$_6$~\cite{Ballet-89}.

In contrast to other pyroxenes the results of total energy calculations 
in the case of Fe-based pyroxenes differ from those obtained in LEIP
formalism. The solution with AFM ordering in the chain was found 
to have lowest total energy for both Fe-based pyroxenes, and
the intrachain exchanges extracted from comparison of AFM and FM total energies are  7~K and 8.5~K for LiFeSi$_2$O$_6$
and NaFeSi$_2$O$_6$ respectively. The difference with the LEIP results appears because of the 
strong polarization of oxygen $2p$-shell. This contribution is not
explicitly (only via hybridization effects) taken into account in the present
LEIP formulation, since both the Green functions 
used in Eq.~\eqref{LEIP} and defined in Eq.~\eqref{Green},
and the potential correction $\Delta_{ij}$ are calculated
only for transition metal $3d-$states. The values of exchange constants obtained 
from the total energies  agree much better with the experimental values of the Curie temperature.

The detailed explanation of the origin of the difference in the results for the exchange constants obtained by LEIP and from the total energies for Fe pyroxenes, is the following. According to the crystal structure of pyroxenes all the oxygens surrounding
every
transition metal ion can be divided on two categories. There are 
two ``side-oxygens'' linking TMO$_6$ octahedra and Si(Ge)O$_4$ tetrahedra,
and four oxygens connecting neighboring octahedra,
as shown in Fig.~\ref{topology}(b). In the case of
AFM intrachain coupling, those four oxygens are situated in the
compensated field and will not have any polarization.
Two other oxygens may be magnetized if the exchange field
is large enough. For the FM chains both types of oxygens will be
polarized. Such a polarization will lead to an addition
magnetic contribution in the total energy calculation.
\begin{center}
\begin{figure}[t!]
 \centering
 \includegraphics[clip=false,width=0.3\textwidth]{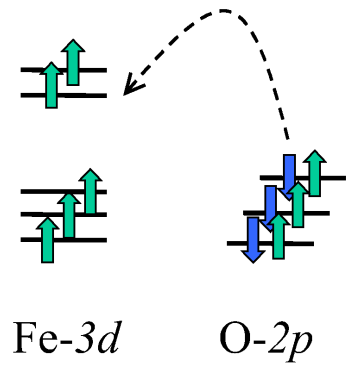}
\caption{\label{polarization}(Color online). 
Sketch illustrating mechanism of the spin-polarization on oxygen 
ions in Fe-based pyroxenes.} 
\end{figure}
\end{center}
\noindent

This effect is especially important and pronounced for the pyroxenes,
based on the late transition metal ions. Those ions have
large magnetic moment and hence could easier magnetize oxygens. Even more important is the other factor: the 
non-zero magnetization of oxygen implies strong
mixing of magnetic $d-$states of TM with one of O$-2p$ orbitals
of the same spin, and therefore it
will be much stronger for TM ions with partially filled $e_g$ sub-shell.
For Fe-based pyroxenes with half-filled $e_g$ shell this effect is 
rather large: the average
magnetic moment on oxygens, lying between TMs,
is 0.17~$\mu_B$ for LiFeSi$_2$O$_6$ and 0.16~$\mu_B$ 
for NaFeSi$_2$O$_6$. The oxygen moments are
directed parallel to those of Fe. It is clearly seen from
Fig.~\ref{polarization}: only hoppings with antiparallel spin 
are possible between O$-2p$ to Fe$-3d$ shells.

Additional polarization of O$-2p$ shell in FM solution will lower 
its total energy. Qualitatively one may consider it as following. 
In the AFM case both spin up and down electrons would hop from oxygen to
different iron ions, while only spin down electrons will be active ones in  FM situation, as
shown in Fig.~\ref{polarization}. Because of the Hund's rule coupling
on oxygen, the latter (FM solution) will gain more energy.

It is easy to estimate the lowering of the FM total energy because of
the oxygen polarization recalling that the gain in magnetic
energy in the band theory equals to
\begin{equation}
\label{ox-contribution}
E_O = - I M^2/4, 
\end{equation}
where $I$ is the Stoner parameter defined as the second derivative of the
exchange-correlation energy with respect to the magnetic moment $M$:
\begin{equation}
I = -2 \partial^2 E_{xc}/ \partial^2 M. 
\end{equation}
The Stoner exchange constant $I$ for the oxygen $2p$-shell was estimated to
be $I=1.6$~eV in Ref.~\onlinecite{Mazin-97,Mazurenko-07} using 
constrained LDA calculations. 

Recalculating the total energy contribution given by Eq.~\eqref{ox-contribution}  
in terms of the exchange constants for Heisenberg Hamiltonian and 
taking into account that there are four polarized oxygens per two formula
(without two ``side-oxygens'') we obtained that the exchange correction
due to the oxygen polarization is 10.7~K and 9.5~K for LiFeSi$_2$O$_6$ and 
NaFeSi$_2$O$_6$ respectively. 
The difference between these values and those number obtained
in LEIP formalism is a little bit larger than the exchange extracted
from the total energy calculations. This seems to be
because we did not account for the changes in the one-electron
energies (relaxations), which tends to compensate the magnetic
corrections given in Eq.~\eqref{ox-contribution}.

Finally, it is important to note that the contribution, which comes 
from the polarization effects on oxygen, is not a unique feature of
pyroxenes. It was recently found to be crucial for the description
of the magnetic interactions in Li$_2$CuO$_4$~\cite{Mazurenko-07}.
Authors of Ref.~\onlinecite{Mazurenko-07} originally interpreted this effect in
the manner of initial ideas of Heisenberg as the exchange between
Wannier functions (constructed out of TM-$d$ and O-$2p$ orbitals) 
due to the Coulomb interaction. It is interesting to note that 
the final expression for
the correction presented in Eq.(22) of Ref.~\onlinecite{Mazurenko-07}
is exactly the same as ours written in Eq.~\eqref{ox-contribution}, if one
uses substitution: $\beta = M(O)/2$.   

\section{Interchain exchange}
In addition to the intrachain exchange interaction we calculated for 
Fe-based pyroxenes the interchain ones. 
It is especially important for the analysis of a possible origin of
multiferroicity in $Fe$-based pyroxenes~\cite{Jodlauk-07}. The results reflect
the general tendency for
interchain exchanges in the whole pyroxenes family. There are two types
of interchain paths, as depicted in Fig.~\ref{topology}(a): some of
TM ions in different chains are connected via two SiO$_4$ (GeO$_4$) tetrahedra ($J_2$),
while the others -- via only one ($J_1$).

First of all, we found that the interchain exchange is much weaker than the intrachain one. Then, we found that for both LiFeSi$_2$O$_6$ and NaFeSi$_2$O$_6$ $J_2$ is,
indeed, approximately two times larger than $J_1$: 
$J^{Li}_1$ = 1.9~K,  $J^{Li}_2$ = 3.4~K, 
$J^{Na}_1$ = 0.8~K,  $J^{Na}_2$ = 1.6~K.
In addition one may see that the change of Li ions to Na leads to a
significant decrease of interchain exchange interaction. It is
connected with larger ionic radii of Na (in 6-fold coordination
$R_{Li^+}$=0.76 $\AA$, while $R_{Na^+}$=1.02 $\AA$).

\section{Conclusions} 
In this paper, using ab-initio LSDA+U calculations, we carried out the detailed 
analysis of electronic structure and 
especially of the exchange interaction in a broad class of quasi-one-dimensional 
magnetic systems --  pyroxenes containing transition metal ions. 
We analyzed the systematics of the exchange interactions and compared it with the 
simple rules following from the perturbation theory.

The exchange interaction due to direct hopping
between $t_{2g}$ orbitals (which we call ``direct exchange interaction'')
dominates in early transition (Ti,V) metal-based pyroxenes. In NaTiSi$_2$O$_6$ 
single $d-$electron localizes on the $zy-$orbital, which 
results in a strong AFM coupling in one out of two Ti-Ti pairs.
This leads to the dimerization of TiO$_6$ chains
and the formation of a spin gap on singlet Ti-Ti dimers.

The nonuniformity of the exchange interaction disappears
in $d^2$ pyroxenes, such as (Li,Na)VSi$_2$O$_6$, where
electrons occupy $zx$ and $zy$ orbitals. This type of orbital
filling is determined by a common structural feature of all pyroxenes
(except Jahn-Teller ones): the presence of two short TM-O bonds
perpendicular to a local $z-$axis (directed along single
180$^{\circ}$ $O-TM-O$ bond connecting two neighboring
TMO$_6$ octahedra).

With an increase of the number of $d-$electrons AFM $t_{2g}-t_{2g}$ direct
exchange interaction is gradually suppressed due to an increase of the
on-cite Coulomb interaction ($U$), and the $t_{2g}-t_{2g}/e_g$ super-exchange 
contributions (most of which are FM)
starts to play a more important role, especially because of the decrease of the 
charge-transfer energy, which enters the corresponding expressions for the exchange 
in Eqs.~(\ref{t2g-t2g-hf}-\ref{diff}) in denominator.
In Cr-based pyroxenes ($d^3$) the AFM $t_{2g}-t_{2g}$ exchange interaction is
nearly compensated by the FM $t_{2g}-e_g$ exchange.
The crystal structure makes a fine-tuning of these contributions,
and NaCrGe$_2$O$_6$ with largest $Cr-Cr$ distance turns out
to be FM.

In NaMnSi$_2$O$_6$, Mn$^{3+}$ is a Jahn-Teller ion. We show that 
the consecutive ``twisting'' geometry of pyroxenes and 
the tendency to gain an elastic energy result in the ferro type of
distortion with the local $z$-axis becoming the longest one.

Iron-based pyroxenes were found to be AFM, with the large oxygen
polarization ($\mu_O \sim 0.2 \mu_B$). This effect 
 is usually ignored in the literature~\cite{Goodenough}. However, we
show that it considerably reduces (by $\sim$ 2/3) the  AFM super-exchange
interaction. The total energy of the FM state is lowered, because
of the gain in Hund's rule coupling in O$-2p$ shell, which
turns out be polarized in this state. 

The interchain exchange coupling was estimated for Fe-based pyroxenes
and found to be small and AFM.

 The presented results give a rather complete description of exchange
 interactions in magnetic pyroxenes, and it may serve as a basis for the further 
 study of magnetism and other properties, including multiferroicity, of this 
 interesting class of materials.

\section{Acknowledgments}
We are grateful to Prof. Dr. G. Redhammer, who took up our call to measure the
crystal structure of NaCrGe$_2$O$_6$, to Dr. N. Binggeli for the help
with pseudopotential code and to Prof. Dr. V. Anisimov and 
Dr. V. Mazurenko for their invaluable help in interpretations of magnetic
calculations for Fe-based pyroxenes.

This work is supported by Dynasty Foundation and International Center for 
Fundamental Physics in Moscow, by INTAS via YS fellowship 05-109-4727, 
by the Russian Ministry of Science and Education together with the Civil
research and development foundation through grant Y4-P-05-15, 
by the Russian president grant for young scientists MK-1184.2007.2,
by the Russian Foundation for Basic Research through RFFI-07-02-00041 and 
06-02-81017,  and by SFB 608.


\begin{thebibliography}{21}
\bibitem{direct-exch} One should
not mix our direct exchange with the standard Heisenberg exchange.
The first one is caused by virtual hoppings of
$d$-electrons to neighboring sites, $J\propto 2t_{ij}^2/U_{dd}$, 
and is not due to an exchange part of the Coulomb interaction, 
$J^{Heis}_{ij} = \int  \phi^{*}_{i} (r) \phi_{j} (r) \phi^{*}_{j} (r') \phi_{i} (r')
/ (|r-r'|) dr dr'$, as the second one.
\bibitem{precision} It is hard to estimate explicitly a precision of
the band structure calculation for the exchange constants computations,
since it depends on many implicit factors such as type of the approximation
and etc. We tried to change some internal parameters of our
calculation scheme as $R_{MT}$, linearization energies and so on
and estimate it to be 2-3~K for $S=3/2$.
\end{thebibliography}
\end{document}